\documentclass[useAMS]{mn2e}

\usepackage{ulem} 
\usepackage{graphicx}
\usepackage{amsmath}
\usepackage{ulem}
\usepackage{rotating}
\usepackage{amssymb}

\def\lapprox{\hbox{\lower .8ex\hbox{$\,\buildrel < \over\sim\,$}}}
\def\gapprox{\hbox{\lower .8ex\hbox{$\,\buildrel > \over\sim\,$}}}

\title[comet NEOWISE]{
Coma morphology and dust emission pattern of comet C/2020 F3 (NEOWISE)
}
\author[F.\,Manzini et al.]{F.\,Manzini$^{1}$\thanks{E-mail: manzini.ff@aruba.it}, V.\,Oldani$^{1}$, P.\,Ochner$^{2,3}$, E.\,Barbotin$^{4}$, L.\,R.\,Bedin$^{2}$, R.\, Behrend$^{5}$,
\newauthor and G.\,Fardelli$^{6}$\\ 
$^{1}$Stazione Astronomica di Sozzago, Cascina Guascona, I-28060 Sozzago (Novara), Italy\\
$^{2}$INAF-Osservatorio Astronomico di Padova, Vicolo dell'Osservatorio 5, I-35122 Padova, Italy\\
$^{3}$Department of Physics and Astronomy-University of Padova, Via F. Marzolo 8, I-35131 Padova, Italy\\
$^{4}$GVO, F-16250 Etriac, France\\
$^{5}$Geneva Observatory, CH-1290 Sauverny, Switzerland\\
$^{6}$Via Oberdan 81A, I-64100 Ascoli Piceno, Italy\\
}

\begin{document} 

\date{Accepted 2021 June 14. Received 2021 June 8; in original form 2021 February 18.}

\pagerange{\pageref{firstpage}--\pageref{lastpage}} \pubyear{201X}

\maketitle
 
\label{firstpage}

\begin{abstract}
The recent close approach of comet C/2020 F3 (NEOWISE) allowed us to
study the morphology of its inner coma.  From the
measurement of the dust ejection velocity on spiral structures
expanding around the nucleus, we estimated a mean deprojected
expansion velocity $V_{\rm d}$ = 1.11$\pm$0.08\,km\,s$^{-1}$. Assuming
that a new shell formed after every rotation of the comet, a rotation
period of 7.8$\pm$0.2 hours was derived. The spin axis orientation was estimated at RA 210$^\circ$$\pm$10$^\circ$, Dec. +35$^\circ\pm10^\circ$.  The coma morphology
appears related to two strong, diametrically opposite emissions
located at mid-latitudes on the nucleus.  A qualitative modelling of
the coma produced consistent results with a wide range of dust
sizes (0.80 to 800\,$\mu$m), with inversely correlated densities
(0.003 to 3.0\,g\,cm$^{-3}$).  Images taken with $Vj$ and $r$-Sloan
filters showed a greater concentration of dust in the first two
shells, and an increasing density of radicals emitting in the $B$ and $V$
band-passes from the third shell outwards.  Striae-like structures in the tail suggest that dust particles have different sizes.
\end{abstract}


%
\section{Introduction}
%
%
Comet C/2020 F3 was discovered on 2020 March 27, in the context of the
NEOWISE program of the NASA's Wide-field Infrared Survey Explorer
(WISE) space telescope. At that time the comet was located 2.0 AU from
the Sun and 1.7 AU from Earth, appearing as an object of 18th
magnitude.  C/2020 F3 reached its perihelion on 2020 July 3, at a
distance of only 0.29 AU from the Sun. Its closest approach to Earth
occurred on 2020 July 23, at a distance of 0.69 AU, when it was at
0.64 AU from the Sun. The comet reached its maximum brightness just
after perihelion at magnitude 0.5, becoming the brightest comet
visible from the Northern hemisphere after comet Hale-Bopp in 1996 and
comet McNaught in 2006.

The extremely favourable conditions of observation and the remarkable
activity shown by this comet allowed us to carry out an in-depth study
of the phenomena occurring in the inner coma.

%

%
\section{OBSERVATIONS AND DATA ANALYSIS} 
%
%

%
%
\subsection{Observing instruments} 
%
%
The CCD images used in this work were collected by our team over
forty-two observing sessions from 2020 July 7 to September 3, with
telescopes ranging in diameter from 0.3 to 1.82\,m. The telescopes
used have a spatial resolution between 0.25 and 0.89 arcsec/pixel; the
physical resolution in the collected images ranges approximately
between 270 and 700\,km/pixel on the sky plane at the distance of the
comet.

The full list of images used for this work is provided in Table\,1.  In
the first column the date and mean time (UT) of the observation are
shown.  In the second column the telescopes used for observations are
listed, together with the relevant IAU-MPC (Minor Planet Center -
www.minorplanetcenter.net/iau/mpc.html) code, which allows to trace
the geographic position of the telescopes on the globe. The GVO
observatory located in Etriac (45.52626°, -0.02444°, France), is
awaiting the IAU-MPC code.  The third column shows the sensors used,
which equip the following CCD cameras: Sbig-ST8 at site C10, QHY367C
at San Giacomo, Moravian G4-16000 at GVO, Moravian G3-16200 at site
A12, Moravian G4-16000 and AFOSC at site 098.  The fourth column shows
the resolution of each telescope in arcsec/pixel.  In the fifth column
the filters used during the observations are indicated: with \textit
{I,B,V,R} we mean filters in the Johnson-Cousins bands, with \textit
{u,r,i,g,z} we mean the respective filters in the Sloan bands.
Finally, the last column indicates the number of exposures within each
observation night and the integration time of each single image.

%
%
\subsection{Data reduction} 
%
%
The data reduction was done using standard IRAF tasks for the images
taken at the Asiago (INAF-OAPD) observatory. The data taken at the GVO
and IAU-MPC A12 sites were reduced with the Prism and AstroArt 7.0
softwares, respectively (\texttt{http://www.prism-astro.com} and
\texttt{http://www.msb-astroart.com)}.

Master bias and dark frames were created nightly, then
subtracted from the raw images. Depending on the availability of
twilight flats, we median combined sky flats frames (at GVO and
IAU-MPC A12) or dome flats frames (at IAU-MPC 098) to create a single
master flat frame for each night. Each single flat frame was reduced
with the corresponding master bias and master dark.  Finally, each
bias- and dark-subtracted original image was divided by the master
flat frame.

%
\begin{table*}
  \caption{
Data used in this work.
    }
  \center
\begin{tabular}{lccccr}
\hline \hline
\texttt{Date (UT)} & Telescope @MPC code & CCD & arcsec/pixel & Filter & N$\times$\texttt{ExpTime} \\ 
\hline
 2020-07-08T02:51$^\ast$ & reflector 0.3m @C10 & KAF-1603me & 1.63 & $I$ & 10$\times$2\,s  \\
 2020-07-10T23:59$^\ast$ & reflector 0.3m @C10 & KAF-1603me & 1.63 & $I$ & 10$\times$6\,s  \\
 2020-07-11T02:04$^\ast$ & refractor 0.1m @S. Giacomo (IT) & IMX 094c & 2.65 & None & 3$\times$60\,s \\
 2020-07-18T16:10 & ob. 85mm @Gobi Desert & Canon ff & 5.8 & None & 40$\times$5\,s \\
 2020-07-21T20:20$^\dagger$ & Copernico 1.82m @098 & E2V CCD42-40 DD & 0.25 & $BVrigz$ & 18$\times$10\,s \\
 2020-07-22T21:20$^\ast$ & reflector 0.5m @GVO (FR) & KAF-16803 & 0.54 & None & 100$\times$20\,s \\
 2020-07-24T20:56$^\ast$ & reflector 0.5m @GVO (FR) & KAF-16803 & 0.54 & None & 45$\times$30\,s \\
 2020-07-25T20:05$^\dagger$ & Schmidt 67/92 @098 & KAF-16803 & 0.87 & $uBVri$ & 34$\times$50\,s \\
 2020-07-26T21:10$^\ast$ & reflector 0.5m @GVO (FR) & KAF-16803 & 0.54 & None & 100$\times$30\,s \\
 2020-07-28T21:00$^\ast$ & reflector 0.5m @GVO (FR) & KAF-16803 & 0.54 & None & 75$\times$30\,s \\
 2020-07-31T20:51$^\ast$ & reflector 0.5m @GVO (FR) & KAF-16803 & 0.54 & None & 30$\times$80\,s \\
 2020-08-03T21:00$^\ast$ & reflector 0.5m @GVO (FR) & KAF-16803 & 0.54 & None & 75$\times$30\,s \\
 2020-08-05T21:00$^\ast$ & reflector 0.5m @GVO (FR) & KAF-16803 & 0.54 & None & 75$\times$30\,s \\
 2020-08-09T19:53$^\dagger$ & Schmidt 67/92 @098 & KAF-16803 & 0.87 & $Br$ & 10$\times$30\,s \\
 2020-08-09T20:07$^\ast$ & reflector 0.4m @A12 & KAF-16200 & 0.86 & $BVR$, None & 80$\times$30\,s \\
 2020-08-18T20:15$^\ast$ & reflector 0.4m @A12 & KAF-16200 & 0.86 & $BVR$, None & 10$\times$180\,s \\
\hline
\multicolumn{5}{l}{$^\ast$ this material is available at our repository:} \\
\multicolumn{5}{l}{\texttt{https://web.oapd.inaf.it/bedin/files/PAPERs\_eMATERIALs/NEOWISE\_C2020F3\_MNRAS/}} \\
\multicolumn{5}{l}{$^\dagger$ this material is available at: \texttt{http://archives.ia2.inaf.it/aao/}} \\

\end{tabular}
\label{data}
\end{table*} 
%

%
%
\subsection{Data analysis: coma} 
%
%
In order to perform a morphological analysis of the comet's inner
coma, where it is easier to see structures due to the outflow of gases
and dust from active sources on the nucleus, all images were processed
with \textsc{Astroart} 7.0 applying several algorithms: median subtraction,
division by 1/$\rho$, radial shift alone with different radii,
Larson-Sekanina with angular shift = 20$^\circ$ and r = 5-pixel radial
shift (Larson \& Sekanina 1984; Samarasinha, Martin \& Larson 2013), all centred
on the optocentre of the comet's nucleus. The comparison between the
results of the above methods showed that the highlighted details were
always the same and in the same place. The latter process is the one
that provided the highest signal, therefore we used it as the basic
algorithm for our analyses, always using the other spatial filters as
a control.

To fully characterise the inner coma features and to verify their
evolution as observed in our images over several observation nights,
we then performed a modelling of the inner coma using a proprietary
software (\textsc{Fase 6}, by P. Pellissier) specifically designed to
reproduce Earth-based observations of the dust coma structures
(Manzini et al. 2016). To this purpose, the first element that is
needed is the position of the comet with respect to the Sun and the
observer. This is easily calculated from the orbital elements of the
comet (available in the JPL/HORIZON website
\texttt{http://ssd.jpl.nasa.gov/horizons.cgi}). The other parameters
needed to run the model are the properties of the dust particles
(albedo, size, density, ejection velocity), and those of the nucleus
(spin axis orientation, rotation period and position of the active
sources). The orientation of the spin axis is the most difficult to
obtain, but it is one of the key parameters for the simulation. It was
estimated from a first analysis of the coma structures orientations
(Vincent, Boehnardt \& Lara 2010), and gradually refined based on the hypothesised
position of the active sources on the nucleus and of their associated
emissions, as well as on the effect of the insolation on the active
sources and of the radiation pressure on the emitted dust. A
trial-and-error procedure was applied until a plausible orientation of
the axis, compatible with a realistic reproduction of the features
visible in the processed images, was achieved.

Finally, to verify the correctness of the estimated position of the
spin axis, we made some models of the nucleus with the software
\textsc{Starry Night PP} v. 5.8.4 (Simulation Curriculum Corp.,
Minnetonka, MN (USA),
\texttt{http://www.simulationcurriculum.com}). The model reproduces
the geometric conditions of observation of the cometary nucleus from
Earth by entering the orbital parameters of the comet and the assumed
direction of the spin axis for each of the dates considered.
%
%
\subsection{Data analysis: tail} 
%
%
To highlight the structure of the tail, particularly the presence of
striae, synchrones and syndines, we applied a gradient operator to
the original wide-field images, which produces an “embossed effect” that
increases the visibility of any existing vertical and horizontal
brightness gradients. The algorithm is given by the formula:
\begin{equation}
I^\prime(x,y)  =  I(x,y) - I(x-dx,y-dy)
\end{equation}
The resulting image is displayed with a linear transfer function.  

We successively compared the findings on the processed images with a
two-dimensional numerical model of the general structure of the tail
obtained by means of the Finson-Probstein equation (Vincent 2014),
using a web-based tool (\textsc{Comet Toolbox} by J.B. Vincent,
\texttt{http://www.comet-toolbox.com}).
%

%
%
\section{Results} 
%
%
%
%
%
%
%

%
%

\begin{figure}
\begin{center}
\includegraphics[width=84mm]{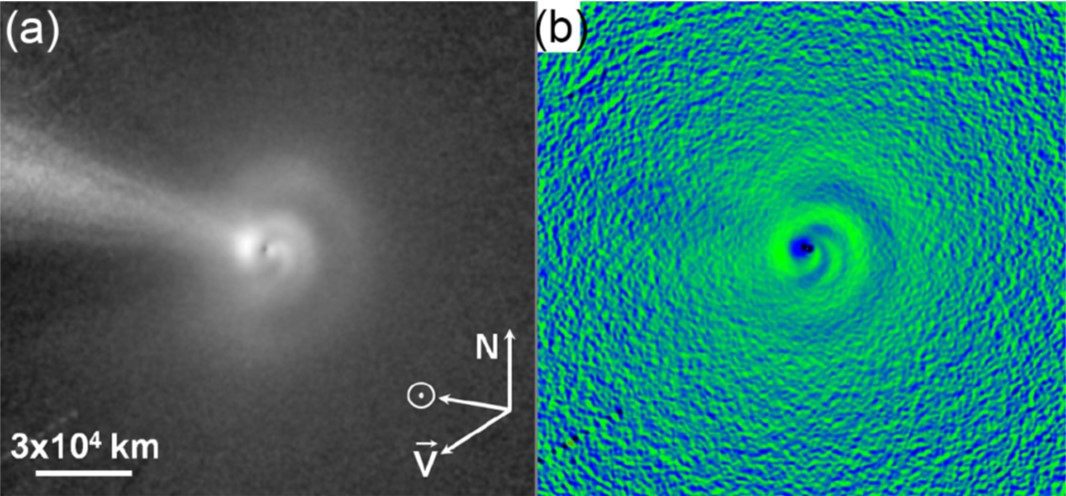}
\caption{
(a) Image of 2020 July 26, processed with a Larson-Sekanina filter with a radial shift to enhance the visibility
of the morphological structures in the inner coma. Nearly full spirals
are visible around the optocentre of comet C/2020 F3, which indicates
that the spin axis was directed almost towards the
Earth. (b) The spirals, originating from two distinct
active sources, are shown more clearly in false colours. North, V and
antisolar vectors are indicated. The angle between the observer and
the comet orbital plane (Plang) was $-$81$^\circ$.
\label{F1}
}
\end{center}
\end{figure}
%

\begin{figure}
\begin{center}
\includegraphics[width=84mm]{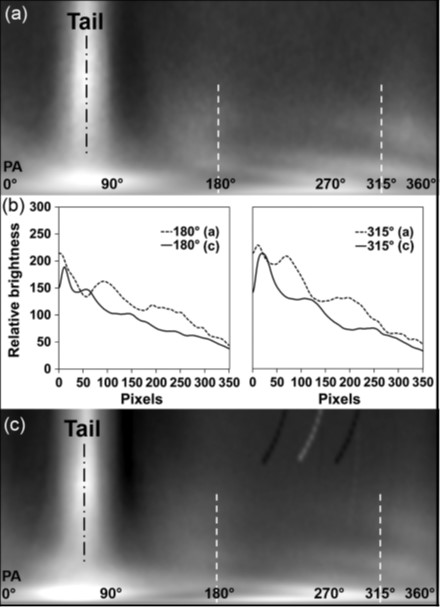}
\caption{
(a and c:) Polar projections, centred on the
optocentre of the false cometary nucleus, of two images of 2020 July
26, processed as in Fig.\,1, taken with 1-hour time interval. The
projections show the development and expansion of the
spirals. (b) The vertical photometric profiles
at PA 180$^\circ$ and 315$^\circ$ provide the precise distance of each
shell from the nucleus (in pixels).
}
\label{F2}
\end{center}
\end{figure}
%
\begin{figure}
\begin{center}
\includegraphics[width=84mm]{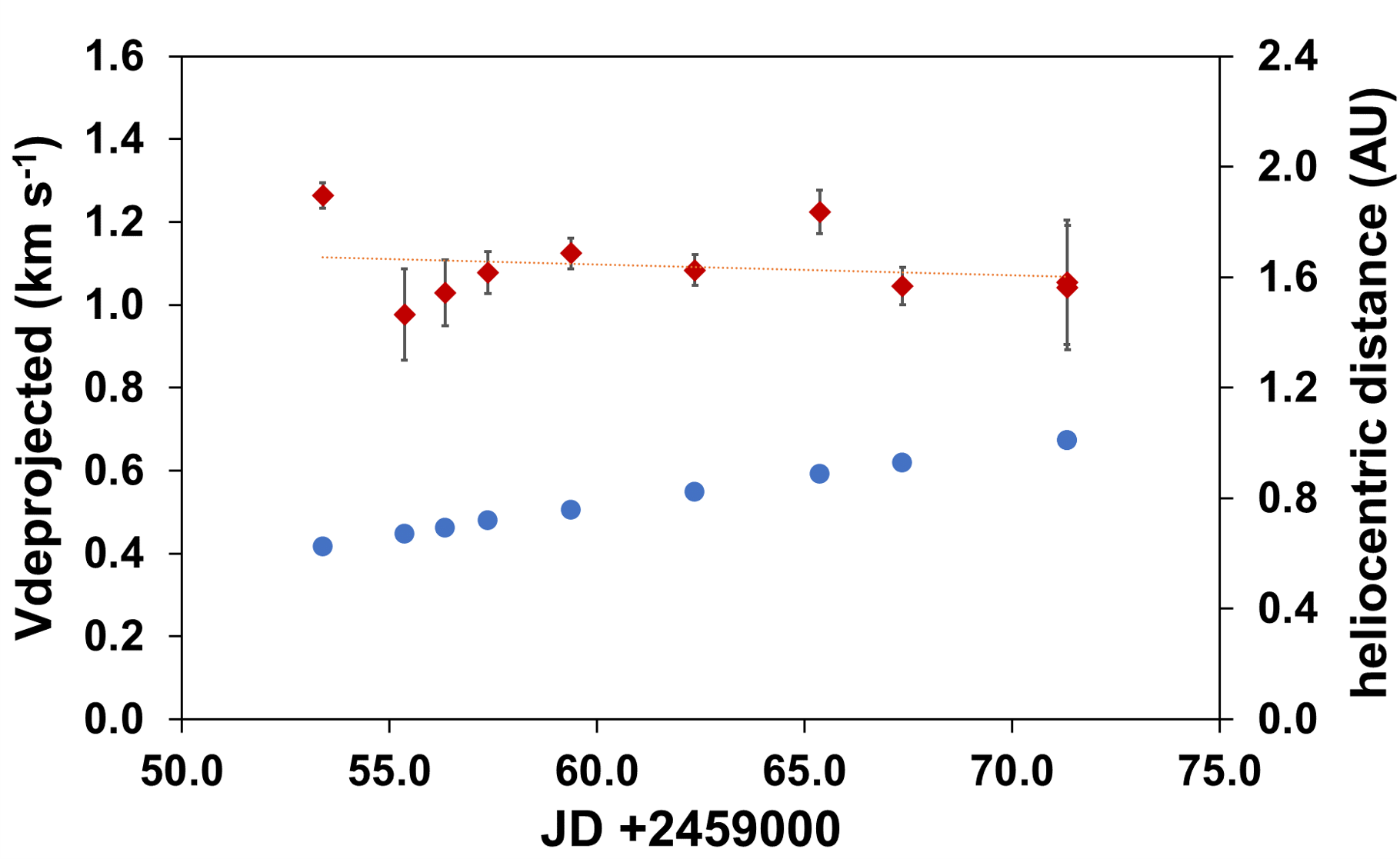}
\caption{
The mean ($\pm$ standard deviation) deprojected expansion velocity
(diamonds) of the spiral-shaped structures measured on ten observing
sessions is shown in relation to the heliocentric distance
(dots). The dotted line shows the linear trend of the expansion
velocity.
\label{F3}
}
\end{center}
\end{figure}

%
\begin{table}
  \caption{ Date, JD, distance from the Sun (r), distance from the
    Earth $(\Delta)$, and measured mean deprojected expansion velocity
    ($V_{\rm d}$ $\pm$ standard deviation) in the observing
    sessions.  } \center
\begin{tabular}{lcccr}
\hline \hline
\texttt{Date (UT)} & JD & r (AU) & $\Delta$ (AU) & $V_{d}$ (\,km\,s$^{-1}$) \\ 
\hline
2020-07-08.10 & 2459038.60 & 0.3256 & 0.9961 & --- \\
2020-07-11.00 & 2459041.50 & 0.3718 & 0.8980 & --- \\
2020-07-11.08 & 2459041.58 & 0.3736 & 0.8948 & --- \\
2020-07-18.67 & 2459049.17 & 0.5322 & 0.7203 & --- \\
2020-07-21.85 & 2459052.35 & 0.6033 & 0.6939 & --- \\
2020-07-22.89 & 2459053.39 & 0.6257 & 0.6919 & 1.264 $\pm$0.030 \\
2020-07-24.86 & 2459055.36 & 0.6703 & 0.6965 & 0.977 $\pm$0.110 \\
2020-07-25.84 & 2459056.34 & 0.6925 & 0.7029 & 1.029 $\pm$0.079 \\
2020-07-26.87 & 2459057.37 & 0.7195 & 0.7119 & 1.078 $\pm$0.051 \\
2020-07-28.87 & 2459059.37 & 0.7585 & 0.7371 & 1.124 $\pm$0.037 \\
2020-07-31.86 & 2459062.36 & 0.8234 & 0.7899 & 1.084 $\pm$0.037 \\
2020-08-03.86 & 2459065.36 & 0.8872 & 0.8568 & 1.224 $\pm$0.053 \\
2020-08-05.86 & 2459067.36 & 0.9292 & 0.9073 & 1.046 $\pm$0.045 \\
2020-08-09.82 & 2459071.32 & 1.0105 & 1.0167 & 1.054 $\pm$0.150 \\
2020-08-09.83 & 2459071.33 & 1.0105 & 1.0167 & 1.041 $\pm$0.150 \\
2020-08-18.80 & 2459080.30 & 1.1889 & 1.2922 & --- \\
\hline
\end {tabular}
\label{data2}
\end{table} 

%

%
\subsection{Coma dust structures} 
%
%
Bow-shaped structures and haloes, which developed between PA
180$^\circ$ and PA 290$^\circ$, subjected to strong radiation
pressure, were visible on the first series of images obtained on 2020
July 10 (comet at r = 0.35 AU). Subsequently, these structures turned
into spiral segments with origin on the cometary nucleus (Manzini et
al. 2020), and became almost full spirals around July 26
(Fig.\,1). By August 15 they had almost vanished and in the
first days of September there was no longer trace of them.

We estimated the dust ejection velocity on two CCD images taken at
least 1 hour apart on ten observing sessions from 2020 July 22 to
August 9 (Table\,2). All images were first transformed into polar maps
centred on the optocentre of the nucleus in order to determine the
position of the dust waves with the highest possible precision. We
measured the shell-to-shell distance in pixels in four different
points at 45$^\circ$ intervals from PA 180$^\circ$ to 315$^\circ$
around the nucleus, averaged the four measures into a single daily
value and then converted this distance $D_{\rm projected}$, projected
on the plane of the sky, into kilometres (Fig.\,2).

The expansion velocity of these structures was calculated by the
simple relation
\begin{equation}
V_{\rm projected} = D_{\rm projected} / \Delta t
\end{equation}
where $\Delta t$ is the time difference in seconds between the two
images on each day. However, since during the considered time-span the
phase angle $\phi$ of the nucleus showed some variation, we also
calculated the deprojected distance for each date to correct the
corresponding deprojected expansion velocities, using:
\begin{equation}
D_{\rm deprojected} = D_{\rm projected} / \sin{\phi}
\end{equation}
Finally, all the resulting daily values were averaged to obtain a
single value for the deprojected expansion velocity $V_{\rm
  deprojected} = 1.11 \pm 0.08$\,km\,s$^{-1}$.

The expansion velocities were essentially steady (or only slightly
decreasing) throughout the analysed period (Fig.\,3). Moreover, the
presence of spiral-shaped structures around the nucleus indicates that
during this time-span the spin axis of the comet was directed almost
towards the Earth (Schleicher \& Farnham 2004). Therefore, the average of
all the measured $V_{\rm deprojected}$ values can be considered as a
realistic estimate of the ejection velocity of dust from the active
areas on the cometary nucleus.

%
\subsection{Rotation period} 
%
%
We made an estimation of the comet's rotation period on the images
taken on 2020 July 26, which showed shells originating from the same
active source with the best resolution (Fig.\,1). Assuming that a new
shell formed after every rotation of the comet, the rotation period of
the nucleus p is simply related to the expansion velocity V of the
shells by the relationship:
\begin{equation}
p = D_{\rm deprojected} \cdot V_{\rm deprojected}^{-1} ~~{\rm (Braunstein~et~al.\,1997)}
\end{equation}
Being the average shell-to-shell relative distance $\sim$30500 km and the deprojected expansion velocity $V_{\rm deprojected}= 1.11$ km\,s$^{-1}$,
a rotation period of 7.8 $\pm$ 0.2 hours could be derived, consistent with the 7.6 hours reported by Drahus et al. (2020).

%
\subsection{Modelling of the inner coma} 
%
%
In order to better characterise the inner coma structures observed on the images and to determine the main properties of the nucleus, we ran a computer model of the inner coma with our proprietary software \textsc{Fase 6}. To this purpose, a number of parameters are needed (as described in Section 2.3).

The previously established parameters, deduced from our measurements were:
\begin{itemize}
    \item for the rotation period, a value of 7.8 hours;
    \item for the dust ejection velocity, the value of $1.11$
      km\,\,$s^{-1}$, estimated as described above, was
      initially entered, although slightly lower values were also
      tested, in line with the observations (Fig.\,3).
\end{itemize}
Other unknown parameters were determined as follows:
    
\begin{itemize}   
     \item for the physical parameters of the dust particles, diametres between 0.80 and 800 $\mu$m, with albedo set at 0.03, were tested in the model. All these particle sizes produced consistent results in the model for inversely correlated values of density between $0.003$ and $3.0 \,$g\,\,$cm^{-3}$, the smallest particles having higher density and the largest particles having lower density. However, the model does not allow to perform a quantitative analysis and to determine the exact dust size distribution.
     \item For the latitude and longitude positions of the active
       source(s) on the nucleus (assumed to be spherical) a
       trial-and-error procedure was followed, moving each
       potentially identified source by steps of five degrees in both
       coordinates, and running the model applying a variable number
       of rotations of the nucleus, until it returned the best
       approximation of the appearance of the features on the
       Earth-based processed CCD images. This approach
       enabled us to relate the complex phenomena observed on the
       images to the presence of at least two
       active sources located at approximately ($\pm$5$^\circ$) the
       following positions on the cometary nucleus:
    \begin{enumerate}
       \item latitude: 40$^\circ$ N; longitude: 0$^\circ$ (arbitrary, taken as reference for the other sources)
       \item latitude: 65$^\circ$ N; at 225$^\circ$ longitude from source (i).
     \end{enumerate}
     
     Considering the geometric conditions of the observation and the
     parameters indicated, these active sources have been identified
     as those responsible for the formation of the spirals expanding
     from the nucleus.
   
   A possible third, additional source was also hypothesised, located
   in this position:

    \begin{enumerate}      
       \item[(iii)] latitude: 80$^\circ$ N; at 45$^\circ$ longitude
         from source (i);
     \end{enumerate}   
   because of its position in latitude on the cometary nucleus, this
   third source produced no spirals or shells, but only an almost
   continuous outflow straight in the direction of the tail at a short
   distance from the nuclear regions.
  
    \item The direction of the spin axis of the nucleus of comet
      C/2020 F3 during the observation period was determined at
      approximately RA 210$^\circ$$\pm$10$^\circ$,
      Dec. +35$^\circ$$\pm$10$^\circ$, with a prograde rotation of the
      nucleus.
\end{itemize}

To best represent the coma structures, the software was set to plot
150 dots at 10-minute intervals over one to six rotations, such as to
reproduce a virtually continuous outflow, and assigning a different
colour to each identified source, in order to distinguish the relevant
dust emissions between each other.

An example of the resulting model, in comparison with the respective
CCD image, is shown in Fig.\,4, where the jets (i) and (ii) are shown
in violet and blue, respectively.

\begin{figure}
\begin{center}
\includegraphics[width=84mm]{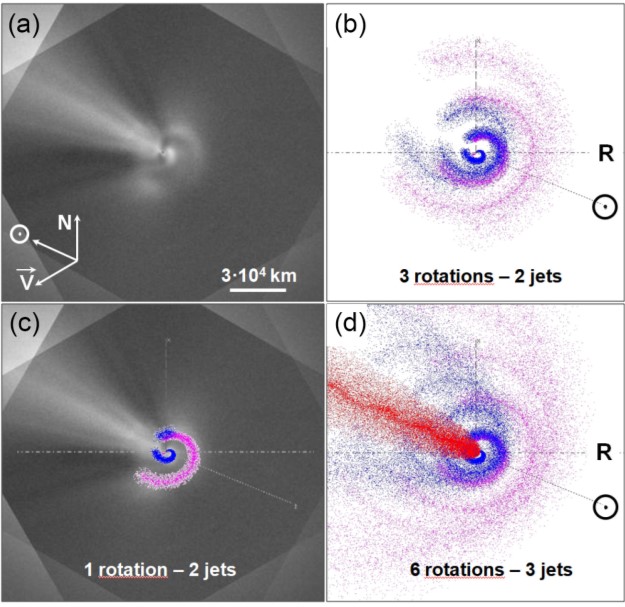}
\caption{
Modelling of the inner coma structures compared with CCD images of
2020 July 24, processed with a Larson-Sekanina filter with a radial
shift(a). The model shows the best agreement with the
observed morphology when three active sources are considered on the
rotating nucleus. In panel (b) two sources and three rotations
are considered to better show the inner coma. (c) The
result of a single rotation with two sources is superimposed to the
CCD image. (d) A third, near-polar jet has been added
to the model and six rotations of the nucleus have been
applied. North, V and solar or antisolar vectors are indicated. The
comet’s spin axis is indicated with R. The orientation of the spin
axis on this date was estimated at PA$\sim$270$^\circ$ with an inclination of
$\sim$75$^\circ$ from the sky plane.
}
\label{F4}
\end{center}
\end{figure}

\begin{figure}
\begin{center}
\includegraphics[width=84mm]{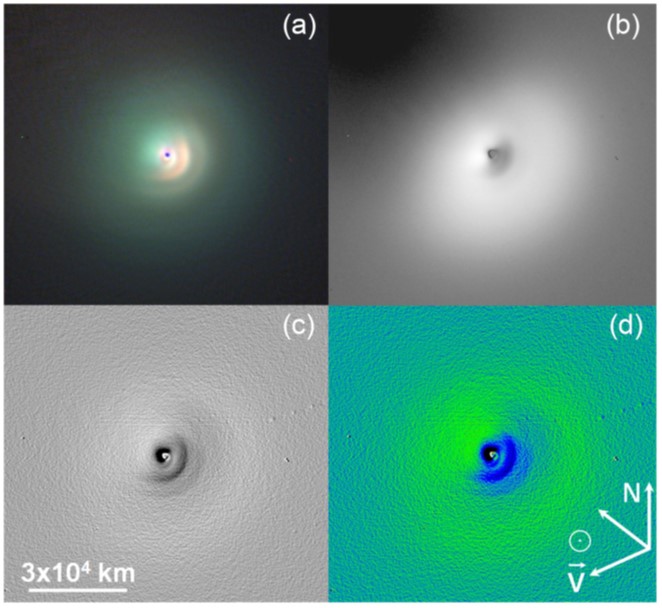}
\caption{
Images of 2020 July 21, taken with the 1.82-m Copernico telescope at
the Asiago Astrophysical Observatory.  (a) Composite
(Bj, Vj, r-Sloan) image calibrated in each colour channel. The dust
component, best shown by the r-Sloan filter, is evident in the
subsolar sector of the first two shells. The coma appears dominated by
the V emissions. The black dot at centre shows the position of the
optocentre of the nucleus.  (b-d) Result of
dividing the Vj image by the r-Sloan image: the dust component
(darker) is spread relatively close to the nucleus, before being
affected by the radiation pressure; the dust tail appears in PA 45$^\circ$.
(c and d) Same image as in panel (b), after applying a
radial shift of 3 pixels to enhance the differences in brightness
around the nucleus, shown in grey shades and false colours.
}
\label{F5}
\end{center}
\end{figure}
\begin{figure}
\begin{center}
\includegraphics[width=84mm]{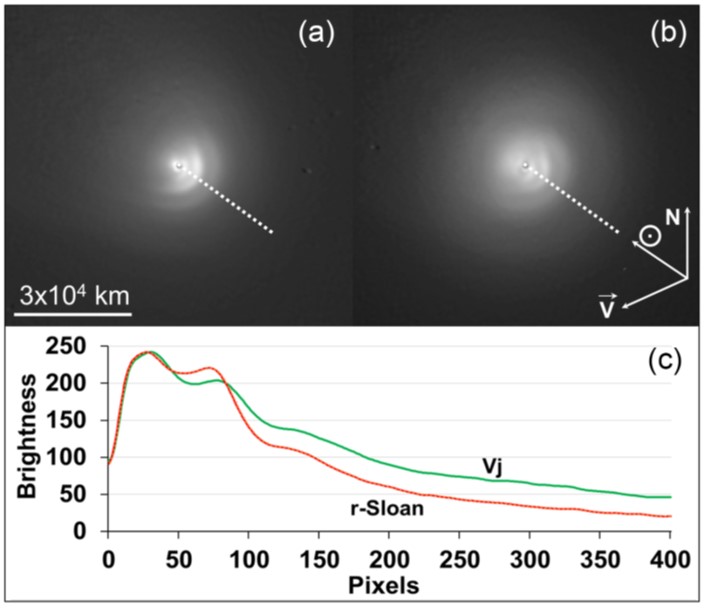}

\caption{
(a and b) Images of 2020 July 21 taken with the
1.82-m Copernico telescope at the Asiago Astrophysical Observatory in
broadband r-Sloan and Vj filters. Resolution: 63 km/pixel.  (c) Brightness profile of the Vj and r-Sloan images traced
for 400 pixels ($\sim$27,000\,km; dotted line in panels (a) and (b)) from
the optocentre (black dot) in the sunward direction. The colour of
each line represents the corresponding filter. In order to make a
direct comparison of the respective slopes, an appropriate vertical
shift was applied to the curves until the peak values corresponding to
the first shell were equalized.
}
\label{F6}
\end{center}
\end{figure}
%
\begin{figure*}
\begin{center}
\includegraphics[width=120mm]{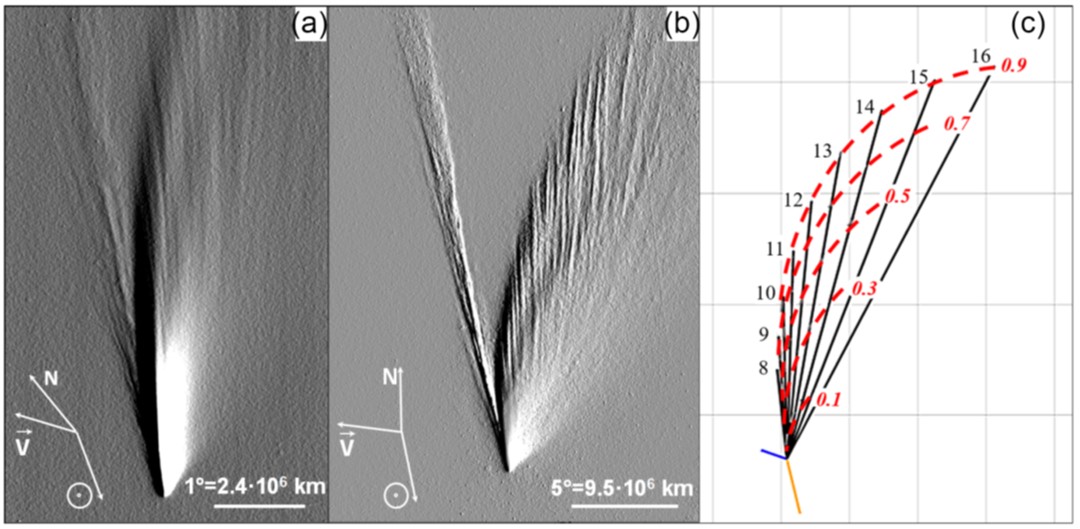}
\caption{
A simple gradient operator was applied to original wide-field images
to increase the visibility of any existing vertical and horizontal
brightness gradients, such as striae, synchrones and
syndines. (a) Image of 2020 July 11.1 UT; UV/IR cut
filter. The angle between the Earth and the comet orbital plane
(Plang) was $-$41$^\circ$. (b) Image of 2020 July 18.7
by L. Zixuan. Plang: $-$62.5$^\circ$. Distances are in km on the sky
plane.
(c) Finson-Probstein diagram plotted on the same date
and with approximately the same scale as the image in
Panel (b) (intervals between vertical gridlines are of
5$^\circ$). The diagram shows the tail geometry with the development
of synchrones (solid lines) over the previous 8 to 16 days and of
syndines (red dashed lines) for $\beta$ values of 0.1 to 0.9. Modified
from: Comet Toolbox by J.B. Vincent,
\texttt{http://www.comet-toolbox.com.}
}
\label{F7}
\end{center}
\end{figure*}
%
\begin{figure}
\begin{center}
\includegraphics[width=84mm]{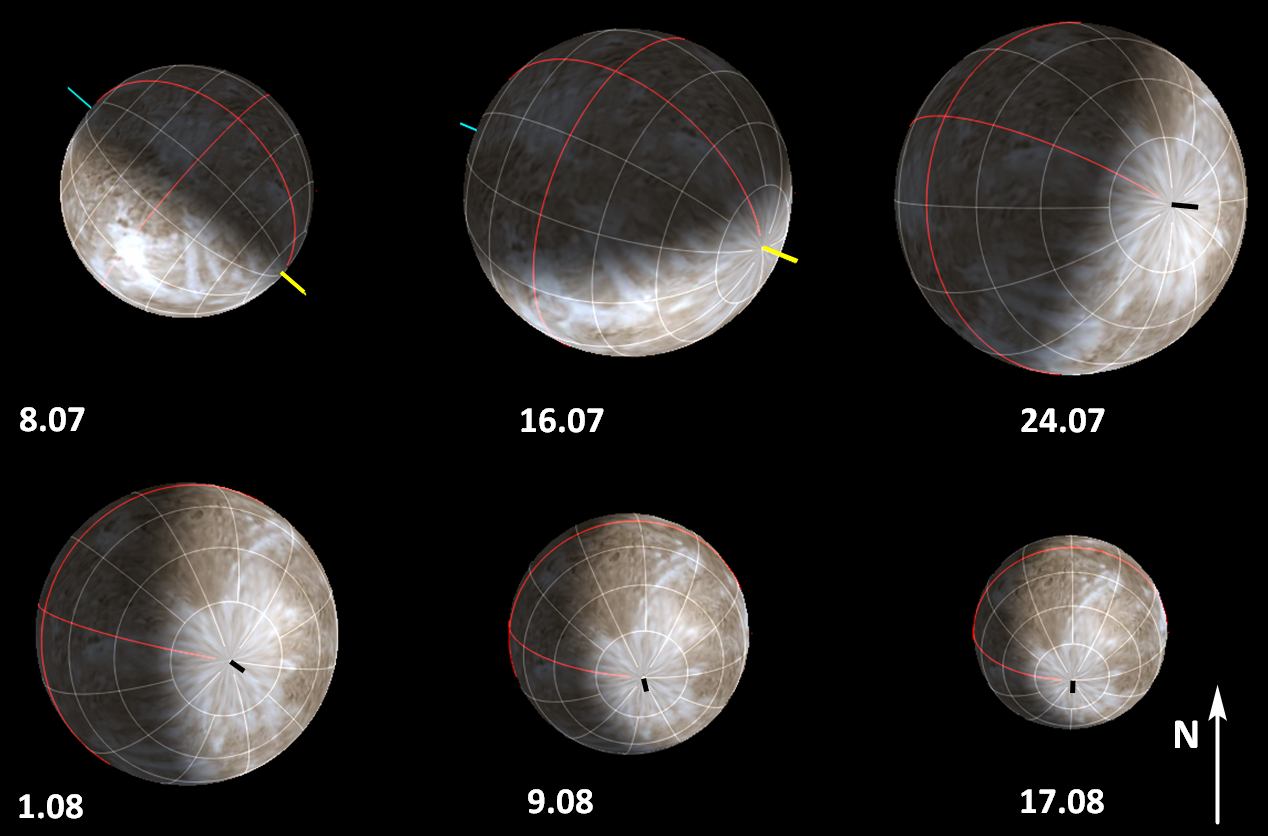}
\caption{
The direction of the spin axis of comet C/2020 F3 as seen from Earth
between 2020 July 8 and August 17 is shown in yellow or black on a
supposedly spherical nucleus, based on its estimated pole position at
RA 210° and Dec. +35°. The phase effect is also shown. The minimum
inclination of the spin axis with respect to the line of sight from
the Earth was reached between late July and mid-August, favouring the
observation of almost complete spirals from emissions deriving from
the active areas on the nucleus.
}
\label{F8}
\end{center}
\end{figure}
%

\begin{figure}
\begin{center}
\includegraphics[width=84mm]{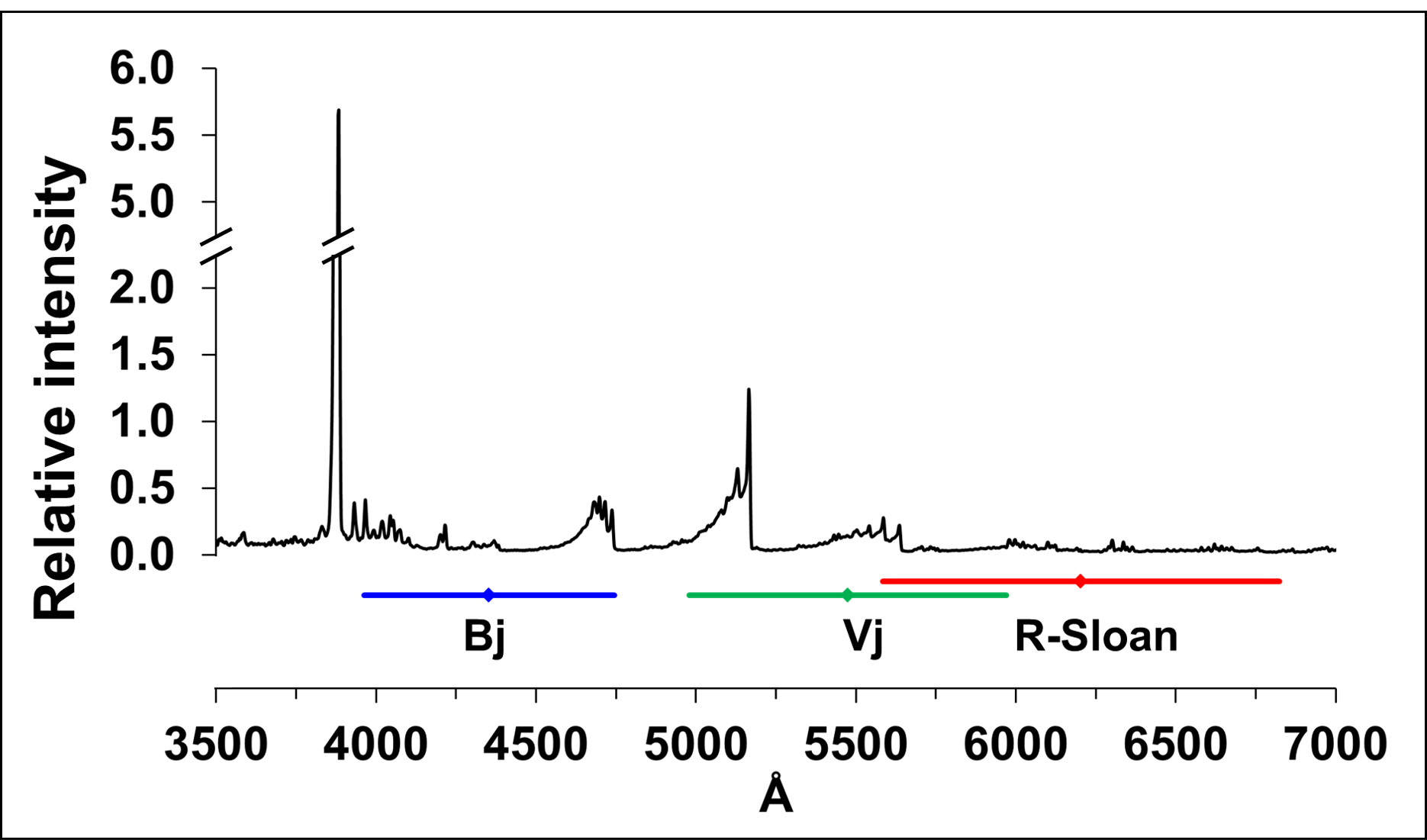}
\caption{
Two spectra were obtained on 2020 August 08 with the 1.22-m Galileo
telescope at the Asiago Astrophysical Observatory with an exposure of
600 s and a resolution of 2.25 {\AA} pixel$^{-1}$.  The two calibrated
spectra were divided by a solar analogue to offset the contribution of
the Sun, then co-added.  The emissions of CN at 3880 {\AA} and the C3
and C2 Swan-bands are very obvious.  Below the spectrum, the
respective central wavelength and FWHM of the Bj, Vj and r-Sloan
filters are shown.  }
\label{F9}
\end{center}
\end{figure}
%


\subsection{Differences between images in Vj and r-Sloan filters} 
%
%
On some of the observation dates, the presence of two or more shells
made it possible, by comparing immediately consecutive images in the
Vj and r-Sloan band, to estimate the distance at which the ratio
between gas and the conveyed dust becomes predominant in favour of the
former or the latter.

Images taken consecutively on 2020 July 21 with the 1.82m Copernico
telescope (INAF-Asiago Astrophysical Observatory) were processed
to assess the difference in the intensity level of the shells between
the data in Vj and in r-Sloan (Fig.\,5).

The same original Vj and r-Sloan images were also processed by means of a
radial shift of 10 pixels to increase the contrast of all those
details showing a gradient of brightness with respect to the
optocentre, followed by a 1/$\rho$ filter to compensate for the radial
fall-off of brightness (Samarasinha, Martin \& Larson 2013). This approach determined a
loss of the photometric data, however the images in the two filters
have been equally processed and the information regarding the
intensity ratio between the gradients of the shell structures was
fully preserved.
Successively, we traced on both images the brightness profiles for 400
pixels in the sunward direction starting from the optocentre, plotted
both curves using an arbitrary scale for the brightness levels and
then compared the peak values corresponding to the three shells for
each curve (Fig.\,6).

The analysis showed that the peak intensity ratio between the first and second shell was:
\begin{itemize}
    \item Vj = 1.20;
    \item r-Sloan = 1.10;
\end{itemize}
and the peak intensity ratio between the first and third shell was: 
\begin{itemize}
    \item Vj = 1.85;
    \item r-Sloan = 2.19;
\end{itemize}
The curve related to the brightness profile of the Vj image showed
also that the peak values corresponding to the shells were slightly
shifted towards the direction of the sun with respect to the r-Sloan
curve.
These data indicate that the contribution of the Vj image was
predominant from the third shell outward, while the contribution of
the r-Sloan image was predominant in the subsolar sector in the first
two shells, which suggests that the dust was more concentrated in that
area, before being affected by the radiation pressure.

The shell expansion over 1-hour time span is showed in a tri-colour
animation (\texttt{C2020F3\_20200724.mp4}, available on-line).


\subsection{Tail structures} 
%
%
The orientation of the tail appeared to be clearly connected to the
anti-solar vector over the entire observation period.

When the comet was between 0.35 AU and 0.55 AU from the Sun (2020 July
10 to July 20), the Earth was far below the comet's orbital plane
(between $-$40$^\circ$ and $-$70$^\circ$) and the nucleus showed a
phase angle of approximately 100$^\circ$. Therefore, the tail was
geometrically projected almost on the plane of the sky, and its
appearance was scarcely affected by the position of the comet in
relation to the Earth. In this period, striae-like structures, as
described by Sekanina \& Farrell (1980) and by Fulle (2004), 
were observed in the tail, probably resulting from dust
fragmentation (Price et al. 2019) or interaction of charged dust with the
solar wind (Fulle et al. 2015). The short-term data we have do not provide
clear evidence of repetitive emissions of material from different
discrete sources on the nucleus, active during the cometary day and
quiescent during the night, so alternately exposed to solar radiation
as the nucleus rotated (Kharchuk \& Korsun 2010).

A deep enhancement processing of the comet’s tail in a wide-field
image of comet C/2020 F3 taken on 2020 July 11 highlighted the
presence of several striae or streamers and two large bands in the
striae probably resulting from the diffusion of dust particles of
different sizes over time (Fig.\,7).

An animation of the motion of the structures and filaments in the tail
is shown in the supplementary material
(\texttt{C2020F3\_20200722.mp4}, available on-line).

In the tail the only significant forces affecting the dusts grain
trajectories are the solar gravity and radiation pressure. These
forces work in opposite directions and depend on the square of the
heliocentric distance. The motion of dust follows the Finson-Probstein
equation 
\begin{equation}
m_{\rm d} \cdot a_{\rm d} = F_{\rm grav} (1-\beta) 
\end{equation}
where $m_{\rm d}$ and $a_{\rm d}$ are the mass and acceleration of the
dust particles, $F_{\rm grav}$ is the solar gravity force, and $\beta$
is the ratio (radiation pressure)/(solar gravity), and is inversely
proportional to the size of the grains for particles larger than 1
$\mu$m (Bohren \& Huffman 1983).

A bi-dimensional numerical model drawn according to the
Finson-Probstein equation for the date of 2020 July 11 fits well with
the direction and length of the tail observed on CCD images taken on
the same date for values of $\beta$ between 0.1 and 0.7.
Assuming a single value of the dust bulk density of
  1.0\,g\,cm$^{-3}$ and a dust scattering efficiency of 1.0, according
  to the Finson-Probstein equation (3) (in: Finson \& Probstein 1968), it
  follows that most of the dust particles had diameters approximately
  between 1.7 and 12\,$\mu$m.
However, it must be considered that this
synchrone-and-syndine model only provides approximate estimates, since
it does not take into account the fact that the dust leaves the coma
with a non-zero ejection speed.

%
\section{Discussion} 
%
%
The applied image processing algorithms are aimed at highlighting
local gradients brightness in the coma, at a distance of a few
thousands kilometres from the surface of the comet’s nucleus, by
reducing the brightness due to the isotropic component of
emissions. This methodology allows to analyse in detail the collimated
structures in the inner coma, however it does not allow to establish
their contribution to the total dust emissions.

The complex morphology that we observed in the inner coma of comet
C/2020 F3 seems to be related to the presence of at least two active
sources located at different latitudes and longitudes on the rotating
nucleus (assumed spherical) and to the position assumed by the spin
axis, with its North pole constantly exposed to sunlight. Both the
geometric conditions of observation from the Earth and the conditions
of insolation of the comet's nucleus due to its orbital motion have
changed during the observation period. As a result, the orientation of
the spin axis showed a slow counter-clockwise rotation, and the
appearance of the morphological features of the inner coma changed
slightly over time (Fig.\,8).

The computer model that we have reconstructed shows the best agreement
with the features observed on the CCD images, in particular the shell
formation, when two almost diametrically opposite sources,
characterized by high-speed emission of dust conveyed by gas, are
placed at mid-latitudes on the cometary nucleus.
Furthermore, the model hypothesised the possible presence of a third
active source located in a near-polar position, which was not
contributing to any of the features of the inner coma, probably due to
a lower emission speed, whereby the dust was more affected by the
solar radiation pressure showing only an almost continuous outflow
straight in the direction of the tail.

The model revealed that the dust outflow from the identified sources
is probably composed of particles in a wide size range, between 0.8
and 800 $\mu$m, compatible with the observations of Rosetta on comet
67P (Fulle et al. 2015; Rotundi et al. 2015; Della Corte et al. 2015) and with the finding
of a significant presence of particles of large size (in the
sub-millimetre range) in the coma by the Giant Metrewave Radio
Telescope (Pal, Manna \& Kale 2021). The model showed an inverse
correlation between the size of the dust particles and their
density. The full dust size distribution is however difficult to
determine since the modelling is purely qualitative. In fact,
modelling the properties of the dust particles is extremely difficult,
as in general only little information about the characteristics of the
dust grains is available and only one dust size at a time can be
entered to run our model.

It should be said that the model only reproduces the spatial
disposition of the dust emitted by discrete sources on the nucleus,
which may only be partially responsible for the formation of the coma
compared to the isotropic release of dust from the sunlit nuclear
surface. In addition, the model assumes a spherical nucleus, whereas a
complex three-dimensional topography of the nucleus may often be
responsible for the development of peculiar coma morphologies (Kramer \& Noack
2015).

The images of the observation session of 2020 July 26 showed the
presence of very obvious spiral shells that were most likely arising
from the same source, but also of other shells that seemed to
originate from a different active area on the cometary nucleus
(Fig.\,1; see also Fig.\,S10 and \texttt{C2020F3\_20200726.mp4},
available on-line).
Indeed, in most of the images taken between mid-July and early August
it appears that the some of the observed shells were due to the
overlap of the emissions from two different sources, as also confirmed
by our modelling (Fig.\,4). This feature supports the presence of two
strongly emitting sources, almost symmetrical to each other, and it
might also be due to a possible slightly different emission speed,
and/or of a variability of the emission speed over time, of the two
sources.

The value of 1.11$\pm$0.08\,km\,\,$s^{-1}$ that we found for
the expansion speed of the observed
structures is higher than those reported for other comets, although it
is not unusual. For example, the expansion speed in comet Hale-Bopp
was measured to range over time from 0.40 to
1.41\,km\,\,$s^{-1}$ (Manzini et al. 2001), and
0.67$\pm$0.07\,km\,\,$s^{-1}$ (Braunstein et al. 1997). Dust speed was
calculated to range from 0.20 to 0.45\,km\,\,$s^{-1}$ in comet
21P/Swift–Tuttle (Jorda, Colas \& Lecacheux 1994; Schulz et al. 1994), comet Levy
1990c (Fulle, Pasian \& Benvenuti 1992), comet P/2004 A1 (Mazzotta Epifani et
al. 2006) and comet Halley’s 1910 passage (Larson \& Sekanina
1984). The Giotto probe measured in situ a neutral gas expansion
velocity of 0.9\,km\,\,$s^{-1}$ (Krankowsky et al. 1986). Comet
8P/Tuttle showed a sunward emission speed of
0.96$\pm$0.03\,km\,\,$s^{-1}$ (Waniak et al. 2009).

The comet's rotation period was estimated at $\sim$7.8 hours; this
value was derived from the measurement of the shell expansion
velocities, which despite was done on images taken in different dates
from those used by Drahus et al. (2020), led to comparable results. The
measured expansion speeds were essentially stable during the
observation period from 2020 July 22 to August 9, with the comet being
at a distance of 0.63 \(<\) r \(<\) 1.01 AU from the Sun. This is
consistent with a steady rotation period of the comet during this
time. Furthermore, the value of the rotation speed that we found was
also confirmed by the modelling; in fact, by applying this period, the
model made a reproduction of the development of the features of the
inner coma very close to those observed on the CCD images according to
the number of rotations applied (Fig.\,4).

The analysis of the images taken with Johnson-Cousins and Sloan
filters made it possible to make a qualitative analysis of the
different contribution of dust and gases to the brightness of the coma
and of the observed structures.  Although the used broadband filters
include both gas and dust emissions, in some cases their use can
provide some hints on the dust-to-gas ratio. Since the r-Sloan
band-pass is centred at $\lambda$=6204\,\AA, with an equivalent width of
$\sim$1240\,\AA, it should be least affected by possibly present gaseous
emissions, as the next strongest bands (i.e., those of C3, C2 and CN)
are located in the blue to green spectral range (Schulz et al. 2003).
Therefore, the comparison of the R-band images with the V- and B-band
images can allow to make a rough qualitative assessment of the
dust-to-gas ratio. In the case of comet NEOWISE F3, this hypothesis is
confirmed by observations: Schleicher, Knight \& Skiff (2020) reported a lower
dust-to-gas ratio than the standard for a comet, and spectroscopic
observations (Krishnakumar et al. 2020; Mugrauer \& Bischoff 2020) reported strong
emissions of CN and C2, confirmed by a spectrographic analysis
performed with the 1.22-m Galileo telescope (INAF-Asiago Astrophysical
Observatory – Padua University) on 2020 August 4 (Fig.\,9).  From these
observations we concluded that the dust was more concentrated in the
subsolar sector in the first two shells, while the gas component was
predominant from the third shell until the edge of the coma (Fig.\,6).

%
\section{Conclusions}
%
From our analyses of the CCD images of comet C/2020 F3, and the modelling of its inner coma features, we found the following:
\begin{enumerate}
    \item the comet is strongly emitting gas and dust, at high
      ejection speed, most likely from two different active sources
      located in almost symmetrical positions at mid-latitudes on the
      nucleus. A possible third source, located in near-polar position
      was also hypothesised.
    \item The comet spin axis is directed at RA 210$^\circ$,
      Dec. +35$^\circ$, with a prograde rotation. The spin axis was
      directed towards the Earth during the whole observation period,
      and spirals and shell-shaped structures, deriving from the
      active sources, become visible in the inner coma.
    \item The rotation period is confirmed at $\sim$7.8 hours. The
      rotation is stable, occurs on a single axis and is not chaotic.
    \item The coma is probably composed of a wide range of dust sizes
      (between 0.80 and 800 $\mu$m), inversely correlated with their
      density values (0.003 to 3.0 \,g\,\,$cm^{-3}$). The exact
      dust size distribution is however difficult to determine.
    \item The dust was more concentrated in the subsolar sector in the
      first two shells, while the gas component was predominant from
      the third shell to the edge of the coma.
    \item The $\beta$ values of the dust particles in the tail
      have been estimated approximately between 0.1 and 0.7 by means
      of a bi-dimensional Finson-Probstein model.

\end{enumerate}

%
%
\section*{Acknowledgments}
This research is based on observations collected at the Copernico and
Schmidt telescopes (Asiago, Italy) of INAF - Osservatorio Astronomico
di Padova.  We would like to thank Lina Tomasella, Jean François
Soulier and Lin Zixuan for sharing important images and data for the
realization of this article, and Gabriele Cremonese for the valuable
advice.

We would like to acknowledge the anonymous reviewer whose helpful
comments significantly improved the quality of this manuscript.
\section*{Data availability}
The images collected with the Copernico and Schmidt telescopes at the
Asiago Observatory are directly downloadable from the INAF
institutional archive interface.  The wide-field image of 2020 July 18
was provided by Lin Zixuan by permission. It will be shared on request
to the corresponding author with permission of Lin Zixuan.  All the
remaining data underlying this article are available at our group
repository at INAF.  The full list of images and the URL of the
corresponding repository are shown in Table\,1.



\label{lastpage}


\end{document}